\documentstyle[12pt,aaspp4,flushrt]{article}
\newcommand{\etal}{\mbox{ et~al.}}
\newcommand{\eg}{\mbox{e.g.}}
\newcommand{\kms}{\mbox{ km~s$^{-1}$}}

\newcommand{\ciii}{CIII$\left.\right]$}

\def\gtorder{\mathrel{\raise.3ex\hbox{$>$}\mkern-14mu
             \lower0.6ex\hbox{$\sim$}}}
\def\ltorder{\mathrel{\raise.3ex\hbox{$<$}\mkern-14mu
             \lower0.6ex\hbox{$\sim$}}}
\def\arcs{\ifmmode {^{\scriptscriptstyle\prime\prime}}
          \else $^{\scriptscriptstyle\prime\prime}$\fi}
\def\arcm{\ifmmode {^{\scriptscriptstyle\prime}}
          \else $^{\scriptscriptstyle\prime}$\fi}
\newdimen\sa  \newdimen\sb
\def\parcs{\sa=.07em \sb=.03em
     \ifmmode $\rlap{.}$^{\scriptscriptstyle\prime\kern -\sb\prime}$\kern -\sa$
     \else \rlap{.}$^{\scriptscriptstyle\prime\kern -\sb\prime}$\kern -\sa\fi}
\def\parcm{\sa=.08em \sb=.03em
     \ifmmode $\rlap{.}\kern\sa$^{\scriptscriptstyle\prime}$\kern-\sb$
     \else \rlap{.}\kern\sa$^{\scriptscriptstyle\prime}$\kern-\sb\fi}
 
\begin{document}
 
\title{MGC~2214+3550: A New Binary Quasar
\footnote{Observations reported here were made with the Multiple Mirror 
Telescope Observatory, which is operated jointly by the University of Arizona
and the Smithsonian Institution.}
}
 
\author{
Jos\'e A. Mu\~noz, Emilio E. Falco, Christopher S. Kochanek \& Joseph Leh\'ar
}
\affil{Harvard-Smithsonian Center for Astrophysics \\
60 Garden Street \\ Cambridge, MA 02138, USA}
\author{Lori K. Herold, Andr\'e\, B. Fletcher \& Bernard F. Burke}
\affil{ Massachusetts Institute of Technology \\
77 Massachusetts Avenue  \\
Cambridge, MA 02139, USA}
\authoremail{jmunoz@cfa.harvard.edu}
%\authoremail{falco@cfa.harvard.edu}
%\authoremail{ckochanek@cfa.harvard.edu}
%\authoremail{jlehar@cfa.harvard.edu}
%\authoremail{lherold@dtm.ciw.edu}
%\authoremail{fletcher@maggie.mit.edu}
%\authoremail{bfburke@athena.mit.edu}
 
\begin{abstract}
We report the discovery of a binary quasar, MGC~2214+3550~A,B,
whose components have similar optical spectra at a redshift $z=0.88$.
The quasars are separated on the sky by $3\farcs0$,
and have a magnitude difference of $\Delta{}m_I=0.5$\,mag. 
The VLA radio map at 3.6\,cm shows a single 47\,mJy radio source 
with a core-jet morphology that is coincident with the brighter 
optical quasar~A. Gravitational lensing is ruled out by 
the lack of radio emission from quasar~B, and the lack of any visible 
galaxies to act as the lens. We conclude that MGC~2214+3550~A and~B are 
physically associated. With a projected separation of $12.7\,h^{-1}$\,kpc 
($\Omega_{0}=1$), MGC~2214+3550~A,B is one of the smallest $z>0.5$ 
binary quasars.  
\end{abstract}

\keywords{Quasars:~individual~(MGC~2214+3550) --- radio galaxies ---
gravitational lensing --- binary quasars }

\section{Introduction}

The first candidate gravitational lens to be discovered, Q~0957+561 
(Walsh et al. 1979), had an image separation ($6\parcs1$) typical of a 
small cluster of galaxies. The cluster was rapidly discovered in optical 
images (Young et al. 1981) and Q~0957+561 is universally believed to
be a gravitational lens.  The third candidate gravitational lens to be
discovered, Q~2345+007
(Weedman \etal\ 1982), also had an image separation ($7\parcs3$) typical
of a small cluster of galaxies.  No such cluster is seen in optical
images, leaving Q~2345+007 as the prototypical ``dark lens'' candidate.
There are now 9  high-redshift quasar pairs similar to Q~2345+007. They
have separations from $3\arcsec$ to $10\arcsec$, identical redshifts 
($\Delta v \ltorder 10^3 \kms$), similar spectra, and no detectable, 
normal object to serve as the lens. For the smaller separation lenses 
and the two wide separation radio lenses
(see Keeton \& Kochanek 1996)\footnote{A current summary of the
lens data is available at \tt{http://cfa-www.harvard.edu/glensdata}}, a
normal galaxy (or cluster) located in the correct position to serve as 
the lens is always seen in HST images of the system, unless the quasar to 
galaxy contrast is too severe to detect a normal galaxy (see Keeton, 
Kochanek \& Falco 1997 for a summary of the optical properties),
so explanations of the problematic pairs must invoke a class of group
to cluster mass objects that have too few stars or too little hot X-ray
emitting gas to be detected locally. However, 
the objects would be $\sim 2$ times more abundant than known massive objects, 
and would contradict most of the generally accepted models of structure
formation (see Kochanek 1995; Wambsganss \etal\ 1995).  Despite periodic 
theoretical attempts (\eg, Jimenez \etal\ 1997) 
and new observations (\eg; Michalitsianos \etal\ 1997; Patnaik \etal\ 1996;
Pello \etal\ 1996; Small, Sargent \& Steidel 1997), little progress 
has been made in confirming or rejecting the wide-separation optical 
quasar pairs as gravitational lenses.

Most of the known lenses were not, in fact, found in surveys of optical quasars
but in imaging surveys of radio sources (\eg; Burke, Leh\'ar \& Conner 1992;
Browne \etal\ 1997; King \& Browne 1996).  
As we discuss in Kochanek, Falco
\& Mu\~noz (1997) a comparison of the optical and radio data yields a 
simple proof that most of the quasar pairs cannot be gravitational lenses.
The 9 problematic doubles are all $O^2$ pairs, in which both quasars are
radio-quiet.  The other detectable 
permutations are $O^2R^2$ pairs in which both
quasars are radio-loud, and $O^2R$ pairs in which one quasar is radio-loud
and the other is radio-quiet.  The most important, disregarded
facts about the wide-separation quasar pairs is the lack of a 
population of radio-loud $O^2R^2$ pairs and the existence
of one $O^2R$ pair.  

The key observational discovery bearing on the ``gravitational 
lens versus binary quasar'' argument was the discovery by 
Djorgovski \etal\ (1987) of the first $O^2R$ pair, PKS~1145--071.  The 
system has an angular separation of $\Delta\theta=4\farcs2$, a small 
magnitude difference of $\Delta m_B=0.83$, and indistinguishable redshifts of 
$z=1.345$.  The lower limit on the radio flux ratio is 500:1, providing 
conclusive evidence that the system is not a gravitational lens.  The
spectral similarities of the two components are not particularly better
or worse than the wide separation quasar pairs or many of the true 
gravitational lenses for that matter.  
Most quasars are radio-quiet, with only $P_R=10\%$ (15\%) 
showing 5~GHz radio fluxes exceeding 50 (10) mJy
(Hooper \etal\ 1997).  Thus, for every
$O^2R$ pair we discovered, we would expect to find $(2 P_R)^{-1} \sim 5$ $O^2$
pairs similar to the claimed dark lenses.      
The very existence of PKS~1145--071
combined with the small fraction of radio-loud quasars essentially rules 
out the gravitational lens hypothesis for most of the $O^2$ quasar pairs. 
The only significant weakness in the chain of inference is the uniqueness of 
the system.

We report here on the discovery of a second $O^2R$ binary quasar, 
MGC~2214+3550, with $z=0.88$, $\Delta m_I=0.5$\,mag and
$\Delta\theta=3\farcs0$. 
We identified the object as a quasar in the course of our ongoing
redshift survey of 177 flat-spectrum radio sources covering the 6\,cm
flux range 50--250\,mJy (Falco, Kochanek \& Mu\~noz 1997). The goal of
our survey is to determine the radio luminosity function for
faint radio sources, to set limits on cosmological models using the
statistics of radio-selected gravitational lenses. A subsample of 108
flat-spectrum sources in the flux range 50--200\,mJy was selected from
the MIT-Green Bank (MG)~II and~III Surveys (Langston \etal\ 1990;
Griffith \etal\ 1990). For each source in our sample, we obtained
$I$~band CCD images for optical identification and
photometry. Finally, we procured low-resolution spectra of the
candidate radio source counterparts with the MMT.  Since the radio
positions are accurate to better than $\sim1$\arcsec, with the errors
dominated by the small systematic offsets between the radio VLA and
optical GSC coordinate reference frames, there was rarely any
ambiguity in the identification of the optical counterparts.  However,
we oriented the spectrograph slit to obtain a spectrum of the next
nearest optical source as a matter of routine. MGC~2214+3550 turned
out to have a visible neighbor within 3\arcsec; when we obtained the
spectra of both objects, we discovered that both were quasars with
indistinguishable redshifts.
In \S2 we describe the optical and radio
data, and in \S3 we discuss whether MGC~2214+3550 A,B is a binary quasar or a
gravitational lens and its consequences.

\section{Observations}

MGC~2214+3550 was initially 
selected for our redshift survey from the single-dish 6\,cm MG III
catalog of Griffith \etal\ (1990). 
An accurate interferometric radio position was obtained
from the MIT Archive of VLA snapshots of the MG~survey radio sources
(MG-VLA: Lawrence\etal\ 1986; Hewitt 1986; Leh\'ar 1991; 
Herold-Jacobson 1996). 
MGC~2214+3550 was observed for 2\,min using the A~configuration at 3.6\,cm. 
The interferometer data were calibrated and mapped using
standard AIPS\footnote{AIPS (Astronomical Image Processing System) is
distributed by the National Radio Astronomy Observatory, which is a facility
of the National Science Foundation operated under cooperative agreement by
Associated Universities, Inc.} procedures, and the flux densities
were scaled to 3C\,286 (Baars et al. 1977).  Three iterations of mapping 
and self-calibration were performed to improve the map quality.
The off-source map rms was $0.169$\,mJy\,beam$^{-1}$,
only $\sim20\%$ higher than the expected thermal noise,
and the FWHM beam size was approximately $0\farcs3$. 
The source has a typical core-jet morphology (see Figure~1),
with a compact core and an associated jet
extending eastwards by $\sim3\arcsec$. 
The peak surface brightness of the core and the jet are
$7.20$\,mJy\,beam$^{-1}$ and $3.60$\,mJy\,beam$^{-1}$, respectively,
and the total VLA interferometer flux density of the source is $47\pm2$\,mJy. 
The VLA radio coordinates for the peak of the compact core are 
$\alpha$=22:14:56.98, $\delta$=35:51:25.8 (J2000.0),
with an estimated astrometric uncertainty of $\sim0\farcs2$
(Lawrence\etal\ 1986). 

After selecting MGC~2214+3550 for our redshift survey, we obtained an $I$~band
image of its optical counterpart with the Fred Lawrence Whipple Observatory
(FLWO) 1.2m telescope; the detector was a Loral 2048$^2$ CCD, with a
Kron-Cousins $I$ filter. The pixel scale of the CCD was $0\farcs315$,
the nominal gain 2.30 electrons/ADU, and the nominal read-out noise 7.0
electrons per pixel.
We  bias-subtracted and flattened the image using standard procedures in
IRAF\footnote{IRAF (Image Reduction 
and Analysis Facility) is distributed by the National Optical Astronomy
Observatories, which are operated by the Association of Universities for
Research in Astronomy, Inc., under contract with the National Science
Foundation.}. 
We used the HST Guide Star Catalog (GSC)
to perform the astrometric identifications. The instrumental magnitudes were
calibrated using a GSC star in our field, with an assumed mean $V-I=1.0$ color
for GSC stars. As a result, our photometry is likely to have absolute
uncertainties of $\sim 0.5$\,mag.
The optical image revealed 2 compact objects, the brighter of which we named A,
and the other B. The pair has a separation
of  $3\farcs02\pm0\farcs01$ in the direction with
PA=13$^{\circ}$ east of north from A (see Figures~1 and~2).
We used the IRAF task ``daophot'' to build an empirical model of the 1\farcs2
FWHM point spread function (PSF), and we found that A and B were unresolved.
After subtracting the PSF, we could not see any significant residual.
Table~1 lists the magnitudes and positions that we obtained for A and B. 

We obtained spectra of A and B with the MMT and the Blue Channel
spectrograph, with a slit of width 1\arcsec\ and a 300 line mm$^{-1}$
grating.  The usable wavelength range is $\sim$3400--8100\,\AA, with a
dispersion of 1.96\,\AA\,pixel$^{-1}$, and an effective resolution of
6.2\,\AA\,~(FWHM).  We took 4 exposures on 3 separate nights (the
journal of observations is in Table~2).  In exposures 1 and 4 we
placed the slit on both components, while in exposures 2 and 3 it was
placed on each component in turn, with the slit perpendicular to the
line A--B, giving us a total of three spectra for each component.
Because detailed spectra were not relevant to the goals of our
redshift survey, and because of our relatively poor observing conditions,
we did not obtain high quality spectra. Nonetheless, all 6 spectra of
the two objects show the same two emission lines, corresponding to
\ciii\ $\lambda$\,1909\,\AA\ and MgII $\lambda$\,2798\,\AA\, at the
same redshift of $z=0.88$.  To improve the signal-to noise ratio
(SNR), we combined the 3 spectra for each of the two QSO components,
and we calibrated the fluxes using the spectrophotometric standard
BD+40.4032, whose spectrum was acquired under photometric
conditions. Unfortunately, we could not acquire standard spectra for
all the nights; thus, our flux calibration is only approximate. The
redshifts for the 2 components, based on the \ciii\
$\lambda$\,1909\,\AA\ and MgII $\lambda$\,2798\,\AA\ emission lines,
are $z_A=0.879\pm0.008$ and $z_B=0.876\pm0.008$. In Table~3 we show
the analysis of the redshifts obtained using each emission line. If
the spectrum of component~A is used as a template in a
cross-correlation with the spectrum of B, a relative velocity of
$v=-148\pm420\,\kms$ is obtained.  The combined MMT spectra for the
A~and B~components are shown in Figure~3.  It is easy to notice the
similarity in the shapes of these spectra, and especially in the
detailed profiles of the \ciii\ and MgII emission lines.  However,
there is a difference between the continua of A and B, with the B
continuum increasing slightly more rapidly than that of A, toward the
blue.  The measured equivalent widths are, for \ciii:
$W_{\lambda}=20\pm7$\,\AA\ ($W_{\lambda}=52\pm30$\,\AA) in A (B); and
for MgII: $W_{\lambda}=85\pm20$\,\AA\ ($W_{\lambda}=54\pm10$\,\AA) in
A (B).  The equivalent widths of lines in one component appear to
differ from those in the other, but the low SNRs in the continua of
our spectra imply that this dissimilarity is marginal. We also
combined all 6 spectra; the emission lines then stand out more
strongly above the continuum, but we could not identify any foreground
absorption features.

We compared the optical and radio data by determining the relative
astrometry of the optical and radio sources, and by determining
an upper limit on the existence of other radio sources in the nearby
field.
We determined the position for optical component~A  with
the program IMWCS\footnote{Originally written at the University of Iowa, and
adapted and amplified by D. Mink at the Smithsonian Astrophysical Observatory.}
to set the world coordinate system in our CCD image. We matched 25 stars in 
the image with the reference catalog USNO-A (Monet 1996).
The final absolute coordinates (see Table~1)
have  a standard error of $\sim0\farcs8$. The coordinate difference between the
optical A~component and the radio source is
$ \Delta\alpha=-0\farcs1\pm0\farcs8$,    
$\Delta\delta=-0\farcs3\pm0\farcs8$.
As an additional test, we determined the coordinates of the A~component 
using 10 GSC stars falling within our CCD frame and
the results were compatible with those given above. Thus, it appears that the
optical counterpart to the radio source is the A~component.  
There is no significant radio emission at the location of component~B,
with an upper limit of $0.17$\,mJy\,beam$^{-1}$, from the rms noise in
the map.  By comparing this to the peak surface brightness of the radio core,
we obtain a lower limit of $\geq42$ on the A/B radio flux ratio. 

Finally, we analyzed the morphologies of the other objects detected
within a 70\arcsec\ radius region around each of the A and B optical quasars,
and found that they are all point sources;
therefore, we could not find any nearby galaxy
or cluster of galaxies that was brighter than $m_I\approx21.5$\,mag.

\section{Discussion}

% LENSING SCENARIO
It is attractive, but difficult, to explain the MGC~2214+3550~A,B
system as two gravitationally lensed images of a single quasar. The
spectra for the two components are similar, with a redshift difference
that is consistent with zero. The angular separation is only
$3\,\arcsec$, easily produced by a galaxy or galaxy-group-sized lens, and the
0.5\,mag image brightness ratio is also typical of gravitational
lensing. The differences in the equivalent widths of \ciii\
$\lambda$\,1909\,\AA\ and MgII $\lambda$\,2798\,\AA, and the
differences between the continua of A and B, are weak evidence against
lensing and are comparable to the differences in the optical
properties of many of the $O^2$ pairs claimed as gravitational lenses.
However, if the radio source is quasar~A (or~B, for that matter), then
the difference between the radio flux ratio ($F_A/F_B\geq42$) and the
optical flux ratio ($F_A/F_B=1.6$) cannot be readily explained by a
lens model.

% FLUX RATIO PROBLEMS
There are three possible explanations for the dissimilar radio and
optical ratios, within the context of a gravitational lens scenario,
but none is likely.  First, extinction of A by $\sim3.5$\,mag is
ruled out by the minimal differences in the spectral continuum
slopes of the two components.  Second, a microlensing fluctuation
making B brighter by at least 3.5 mag is unlikely.  Even for true
point sources we expect an rms magnitude fluctuation of only 1\,mag
(Witt, Mao \& Schechter 1995), and the empirical evidence shows that
the observed microlensing fluctuations are considerably smaller (see
Corrigan \etal\ 1991; Houde \&~Racine 1994), implying that the quasar
source sizes are insufficiently point-like for the images to exhibit
the maximum fluctuation.  With such a large microlensing effect we
would also expect larger differences in the equivalent widths of the
emission lines (\eg, Schneider \& Wambsganss 1990).  
Third, strong differential variations in the
optical and radio fluxes, combined with a suitable time delay
between the two quasar images, may be able to produce the observed
ratios.  Our existing data cannot eliminate this hypothesis, since
we lack optical and radio time series, but it is unlikely.

The flux ratio differences are not a problem if our astrometric
identification is incorrect. In this case, it may be possible that
the radio source is actually
associated with the lens galaxy, so that it lies between the A and B images
of a background radio-quiet quasar.  Such a registration 
is improbable because
there are only about $10^{5}$ AGN over the entire sky with 6\,cm
total flux density above 30\,mJy (Gregory\etal\ 1996; Griffith \&
Wright 1993).  With $\sim 10^{11}$ galaxies in the observable
universe, the probability of any lens galaxy being sufficiently
radio-bright is 1 in $\sim 10^6$. 

% THE KILLER PROBLEM
The last serious problem with the lensing interpretation is
that no lensing material can be seen optically.  The mass to produce
the $3\arcsec$ image separation requires at least an $L_*$ galaxy or
a group of galaxies. An $L_*$ galaxy member of a lensing group should be
easily detected in
our images, since even at the higher than optimal lens redshift of
$z=0.5$ an $L_*$ galaxy would have $m_I \sim 18$.
An anomalously faint lens is
very unlikely, given the presence of lenses with the expected
optical fluxes in all other convincing lens systems (see Keeton,
Kochanek, \& Falco 1997), and given that MGC~2214+3550 would have the
lowest source redshift of any known lensed system.

% UNASSOCIATED QUASARS
We conclude that the MGC~2214+3550~A,B system is a binary quasar with a
projected separation of $12.7\,h^{-1}$\,kpc (for $\Omega=1$), making
it one of the smallest projected separation quasar binaries.  
The velocity differences, if interpreted as due to the Hubble flow,
correspond to a line-of-sight separation of $\sim 0.5 h^{-1}$ Mpc 
($\Omega=1$).  However, the uncertainties in the velocity difference
are so large that a far better estimate can be obtained from the
density associated with the correlation function.
If the quasar-quasar density is $n \propto r^{-1.8}$, 90\% of the objects
have line-of-sight separations smaller than approximately 10 times
the projected separation, or about $130 h^{-1} $ kpc. 
The existence of MGC~2214+3350~A,B proves that PKS~1145--071 was 
not a statistical fluke, and that we really are seeing the
number of $O^2R $ objects expected if most or all the large separation
quasar pairs are binary quasars rather than gravitational lenses.
Adding the absence of wide separation $O^2R^2$ radio pairs,
we believe that the combination of the optical and radio data 
conclusively proves that a fraction $\sim 10^{-3}$ of bright, 
high redshift quasars are members of binary quasar systems. 
In Kochanek \etal\ (1997) we quantify the case against the lens 
interpretation in greater detail, and provide a simple physical 
argument for the excess of binary quasars over that predicted from 
the quasar-quasar correlation function (see Djorgovski 1991) based 
on the physics of galaxy mergers.  

\acknowledgments
We are grateful to A.~Milone for her help and perseverance 
during our MMT observations. 
We also thank A.~Vikhlinin for a FLWO 1.2m image,
L.~Macri for an MMT spectrum, and C.~Keeton for theoretical magnitude estimates.
Our research was supported by the Smithsonian Institution. 
JAM is supported by a postdoctoral fellowship from the Ministerio de
Educaci\'on y Cultura, Spain. 
CSK is supported by NSF grant AST-9401722 and NASA ATP grant
NAG5-4062. JL acknowledges support of NSF grant AST93-03527.

\newpage

%%%%%%%%%%%%%%%%%%%  TABLES %%%%%%%%%%%%%%%%%%%%%%%%%%%%

\begin{deluxetable}{cccc}
\tablecaption{ Optical astrometry and photometry for MGC~2214+3550 A,B }
\tablehead{
Object         & $\alpha$ (J2000) & $\delta$ (J2000) & $m_I$ }
\startdata
 A & 22:14:56.97 $\pm0\farcs8$ & 35:51:25.5 $\pm0\farcs8$ & $18.80\pm0.08$ \nl
 B & $\Delta\alpha_{B-A}=0\farcs82\pm0\farcs01$ & 
     $\Delta\delta_{B-A}=2\farcs94\pm0\farcs01$ & $19.30\pm0.08$ \nl
%L & $\Delta\alpha_{L-A}=1\farcs928$ & $\Delta\delta_{L-A}=1\farcs148$ &$\sim$22 \nl
\tableline
\enddata
\label{tab1}
\end{deluxetable}

\begin{deluxetable}{llccccrc}
%\tablewidth{15.9cm}
\tablecaption{ Journal of MMT observations for MGC~2214+3550 A,B }
\tablehead{
Exp.& \multicolumn{1}{c}{Object}  &
Central $\lambda$& 
\multicolumn{1}{c}{Exp.} & 
\multicolumn{1}{c}{Date}&  
\multicolumn{1}{c}{Air Mass} & 
\multicolumn{1}{c}{P.A.} & 
\multicolumn{1}{c}{Seeing}\nl
No      &         & 
\multicolumn{1}{c}{(\AA)} &  
\multicolumn{1}{c}{(s)}    & \multicolumn{1}{c}{dd/mm/yy}  &       &
\multicolumn{1}{c}{$^{\circ}$ E of N} &
\multicolumn{1}{c}{(arcsec)}
}
\startdata
1   & MGC~2214+3550    &   6000  & 2700 & 08/09/96 &  1.031 & -176.9 & 0.9 \nl
2   & MGC~2214+3550 A  &   6000  & 1769 & 05/07/97 &  1.002 &  -90.9 & 1.1 \nl
3   & MGC~2214+3550 B  &   6000  & 2700 & 05/07/97 &  1.022 &  -83.9 & 1.3 \nl
4   & MGC~2214+3550    &   6000  & 3600 & 09/07/97 &  1.055 &   11.9 & 2.1 \nl
\tableline
\enddata
\label{tab2}
\end{deluxetable}

\begin{deluxetable}{cccccc}
\tablecaption{ Redshift analysis for MGC~2214+3550 A,B }
\tablehead{Object & \multicolumn{2}{c}{\ciii\ $\lambda$\,1909\,\AA}
 & \multicolumn{2}{c}{MgII $\lambda$\,2798\,\AA} & 
Redshift \nl
 & \multicolumn{2}{l}{------------------------------} & 
   \multicolumn{2}{l}{------------------------------} & \nl
  &  $\lambda_{obs}$(\AA) & z & $\lambda_{obs}$(\AA) & z & $<z>$ }
\startdata
A & 3576 & 0.873 & 5272 & 0.884 & 0.879$\pm$0.008 \nl
B & 3570 & 0.870 & 5264 & 0.882 & 0.876$\pm$0.008 \nl
\tableline
\enddata
\label{tab3}
\end{deluxetable}

%%%%%%%%%%%%%%%%%%%%%  FIGURES  %%%%%%%%%%%%%%%%%%%%%%%%%%%%%%%%
\newpage

\begin{figure}[ht]
{\epsfxsize=15cm \epsfbox{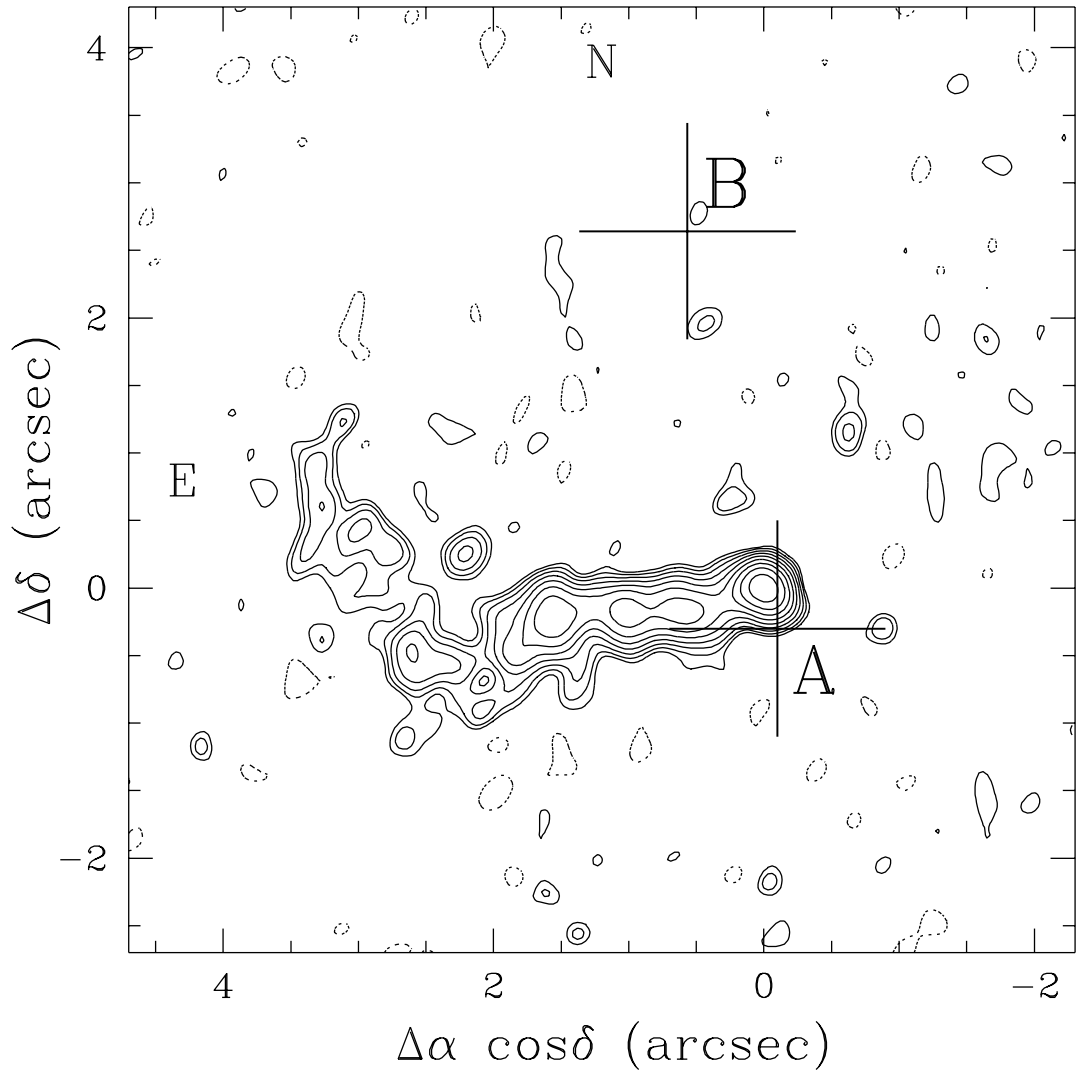}}
\figcaption{Radio total intensity 3.6\,cm contour map of MGC~2214+3550.
North is at the top, and east to the left. Coordinate offsets are given 
relative to the radio core component. 
The positions of the optical quasars A and B are marked with vertical crosses,
the sizes of which indicate the $0\farcs8$ uncertainty in our optical
astrometry. 
The positive (negative) radio flux density is shown as solid (dotted) contours
which increase by factors of $\sqrt{2}$ from twice the off-source map rms level 
of 0.169\,mJy\,beam$^{-1}$; the FWHM beam is $0\farcs285\times0\farcs267$ 
with the major axis oriented at PA=$-80^{\circ}$. 
}
\end{figure}

\begin{figure}[ht]
{\epsfxsize=15cm \epsfbox{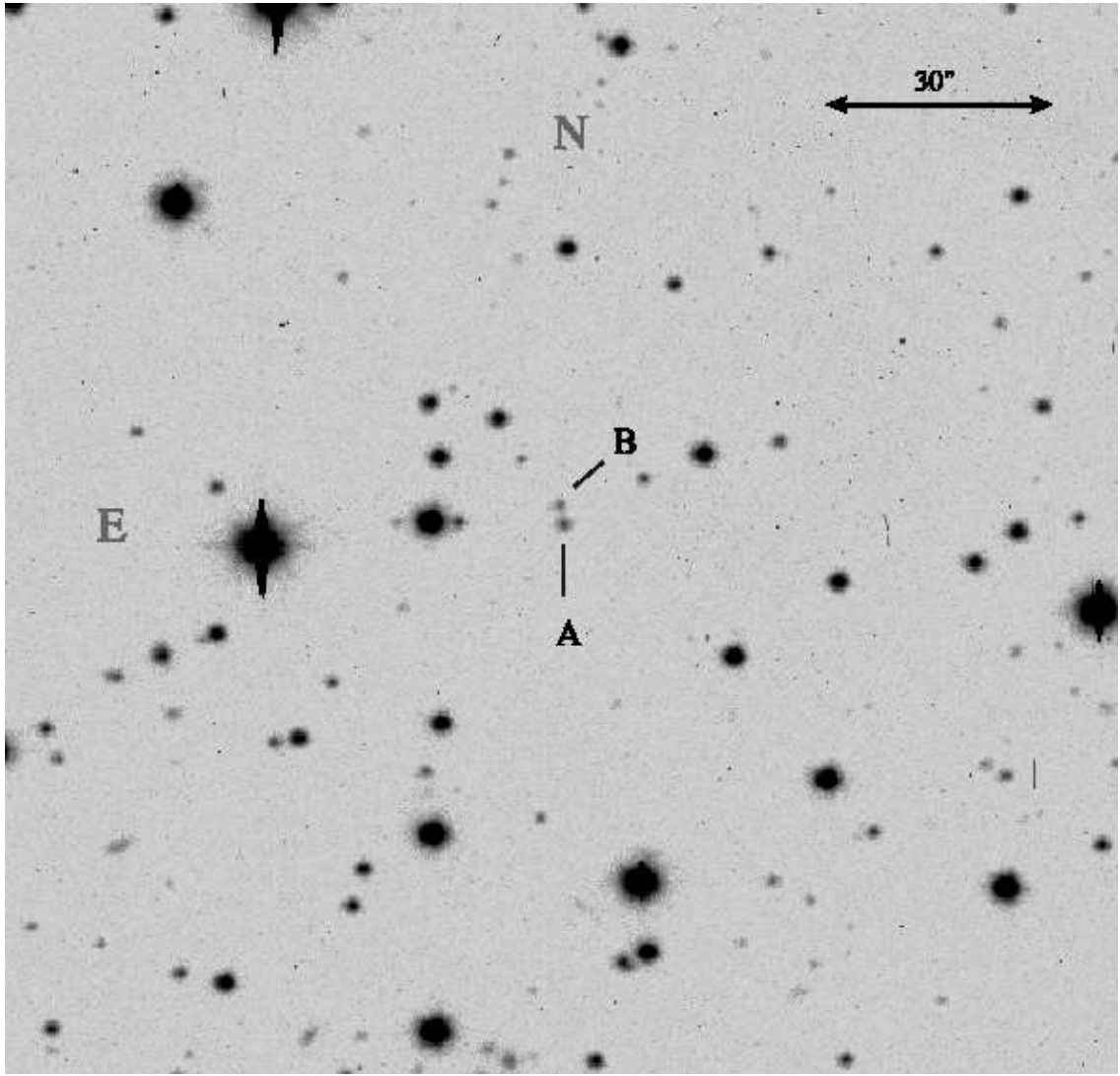}}
\figcaption{Optical $I$~band $2\farcm5 \times 2\farcm5$ CCD image of the
optical field in the direction of the radio source MGC~2214+3550, obtained
with the FLWO 1.2m telescope. The two quasars are indicated by A and B, near
the center of the frame. North is towards the top of the frame, East is to
the left, and the angular scale is shown in the top right corner.
\label{fig2}
}
\end{figure}

\begin{figure}[ht]
{\epsfxsize=15cm \epsfbox{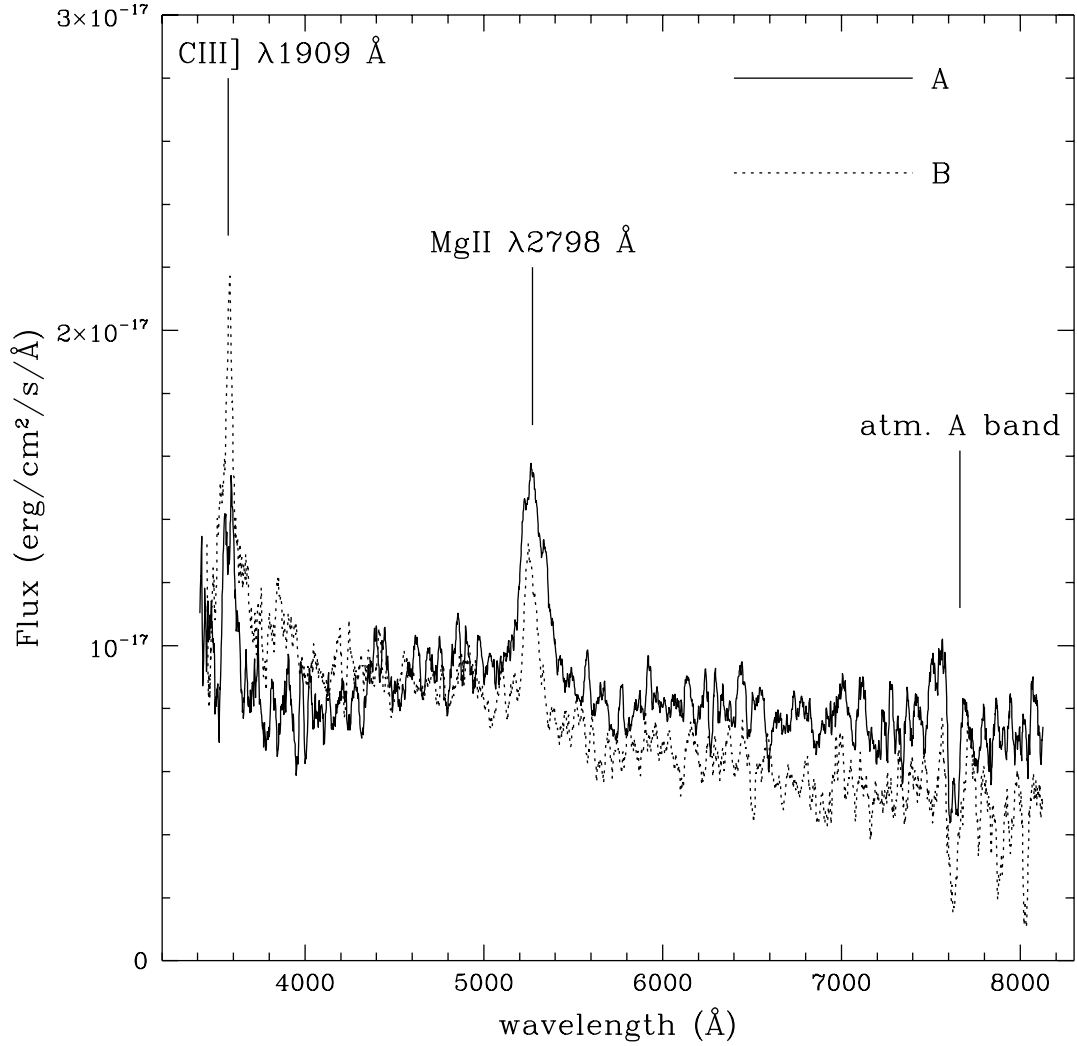}}
\figcaption{Calibrated spectra of the quasar pair in the direction of
MGC~2214+3550, showing the spectrum of each of two QSO components A~\&~B.
The solid (dashed) line corresponds to the A~(B) component. The abscissa shows
observed wavelengths. Prominent emission lines, as well as the atmospheric
absorption A~band, are labeled.
\label{fig3}
}
\end{figure}
\end{document}